\documentclass[12pt]{article}

\usepackage[english]{babel}

\usepackage[letterpaper,top=2cm,bottom=2cm,left=3cm,right=3cm,marginparwidth=1.75cm]{geometry}

\usepackage{amsmath}
\usepackage{amssymb}
\usepackage{amsthm}
\usepackage{graphicx}
\usepackage[colorlinks=true, allcolors=blue]{hyperref}
\usepackage{longtable}
\usepackage[table]{xcolor}
\usepackage{fancyvrb}
\usepackage{tikz}
\usetikzlibrary{automata,arrows,positioning,calc}
\usepackage{soul}
\usepackage[rflt]{floatflt}

\sethlcolor{red!30}

\newtheorem{theorem}{Theorem}

\theoremstyle{definition}

\newtheorem{example}[theorem]{Example}

\title{Rewinding the byte trail of the White Whale}
\author{Boris Alexeev \and Dustin G.\ Mixon\thanks{Department of Mathematics, The Ohio State University, Columbus, Ohio, USA} \thanks{Translational Data Analytics Institute, The Ohio State University, Columbus, Ohio, USA}}
\date{}

\begin{document}
\maketitle

\begin{abstract}
Motivated by a popular code golf challenge, we review some key ideas from information theory and discuss how to efficiently compress a streaming file with an acceptable error rate.
\end{abstract}

\let\thefootnote\relax\footnotetext{DGM was supported by NSF DMS 2220304.}

\vspace{16pt}

\begin{floatingfigure}{0.33\textwidth}
\centering
\includegraphics[width=0.33\textwidth]{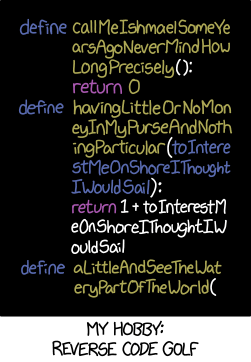}
\end{floatingfigure}
While hoards of normies go golfing on the weekends, there's an underground community of hackers, puzzlers, and codesmiths who take the tee-rific hobby in a pun-derful direction. 
The objective of \textit{code golf} is to write the shortest possible program that performs a given task, i.e., to write a suitable program using the fewest key \textit{strokes}. 
(Ba dum tss!) 
One online haunt for this community is the \textit{Code Golf and Coding Challenges Stack Exchange}, where Nathaniel Virgo once posted a whale of a puzzle called ``Write Moby Dick, approximately''~\cite{Virgo:online}. 
The task here is to write a program that receives the text of Herman Melville's \textit{Moby-Dick; or, The Whale}~\cite{Melville:51} one character at a time, and guesses the next character at each step. 
Unlike most code golf challenges, this one features a twist in how programs are scored: 
The best submission is the one that minimizes
\[
2L+E,
\]
where $L$ is the length of the program in bytes and $E$ is the number of characters it guesses incorrectly. 
The challenge was a massive hit in the community, drawing hundreds of upvotes and even inspiring an xkcd comic~\cite{Munroe:18}.
In this article, we introduce some of the key ideas from \textit{information theory}~\cite{CoverT:99} that lurk behind the current best solution to this challenge (which happens to be due to the authors).

\section*{Primer: A few simpler variations}

To approach this problem, a mathematician's first instinct might be to take a page from George P\'{o}lya's ``How to Solve It''~\cite{Polya:45} by considering variations of the problem. 
For example, instead of \textit{Moby-Dick}, we can take the target document to be any string, perhaps one that's generated by a random process. 
Also, instead of minimizing $2L+E$, one could try to minimize some other linear combination of these variables. 
Let's consider some simple variations.

\begin{example}
\label{ex.1}
Suppose the target document is a string of random bytes, and the goal is to minimize
\[
L+\infty\cdot E.
\]
To achieve a finite score, we seek lossless compression (meaning there are no errors in the reconstruction). 
Interestingly, most strings are not compressible as a consequence of the pigeonhole principle: 
First, the number of byte strings of length exactly $n$ is $256^n$, since a single byte has $2^8=256$ possible values. 
Next, the number of byte strings whose length is strictly less than $n$ is 
\[
\sum_{k=0}^{n-1}256^k
=\frac{256^n-1}{256-1}.
\]
Thus, a compression algorithm can only compress less than 0.4\% of byte strings. 
In some sense (namely, in the sense of \textit{Kolmogorov complexity}), most strings are random. 
Since most target documents cannot be compressed, we can expect the best program to take the form ``output the following string.'' 
For an $n$-byte target document, this achieves a typically optimal score of $n$ (plus the length of the instructions ``output the following string'').
\end{example}

\begin{example}
\label{ex.2}
Next, we consider a target document that’s ``less random'' in some sense: a random string of $\mathtt{E}$s and $\mathtt{T}$s. 
To achieve lossless compression, we can encode a batch of 8 consecutive characters as a byte and then apply our solution from the previous example. 
In particular, the best program is ``output the string encoded by the following hints file,'' but now $L$ is 8 times shorter than the target document since there are only 2 possibilities for each character, whereas each byte in our encoded hints file has $2^8$ possibilities. 
In the parlance of information theory, the size of the hints file equals the total \textit{entropy} of the target document.
\end{example}

\begin{example}
\label{ex.3}
As another baby step towards realism, we now allow for more characters that occur with different frequencies. 
Suppose the target document is a string of $\mathtt{E}$s, $\mathtt{T}$s, and $\mathtt{A}$s, where the characters are independently drawn at random from the following distribution:
\[
\left\{\begin{array}{cl}\mathtt{E}&\text{with probability }49\%\\\mathtt{T}&\text{with probability }49\%\\\mathtt{A}&\text{with probability }\phantom{0}2\%\end{array}\right..
\]
How can we minimize $L+\infty\cdot E$ in this setting? 
Intuitively, $\mathtt{E}$s and $\mathtt{T}$s are the ``usual characters'' here, so we should expect each of these to cost about one bit in our hints file. 
Meanwhile, whenever an $\mathtt{A}$ appears, it comes with some surprise, and the cost should be proportional to some quantification of this surprise. 
In information theory, the \textit{surprise} of a probability-$p$ event is equal to $\log_2(\frac{1}{p})$ bits. 
For example, a random byte equals $01011000$ with probability $p=2^{-8}$, which has surprise 8 bits, and therefore costs 8 bits (i.e., a byte) to encode. 
Meanwhile, the surprises of our characters are
\begin{align*}
\mathrm{surprise}(\mathtt{E})
&=\log_2(1/49\%)\text{ bits}\approx1.0291\text{ bits},\\
\mathrm{surprise}(\mathtt{T})
&=\log_2(1/49\%)\text{ bits}\approx1.0291\text{ bits},\\
\mathrm{surprise}(\mathtt{A})
&=\log_2(1/\phantom{0}2\%)\text{ bits}\approx5.6439\text{ bits}.
\end{align*}
The average surprise of a random character is its entropy:
\[
(1.0291\text{ bits})\cdot 49\% 
+ (1.0291\text{ bits})\cdot 49\%
+(5.6439\text{ bits})\cdot2\%
\approx1.1214\text{ bits}.
\]
As such, we should be able to encode a string consisting of independent realizations of this random character at a price of about 1.1214 bits per character on average. 
One may approach this with \textit{arithmetic coding}. 
This is an exciting topic that we encourage the reader to learn more about, but one interpretation is as follows: 
Partition the unit interval into subintervals whose lengths equal the probabilities of all possible strings, and then encode a given string using a point in the corresponding subinterval with the shortest binary expansion. 
Conveniently, the number of bits in this binary expansion approximates the surprise of the entire string, and so this encoding is essentially optimal.
\end{example}

\begin{example}
\label{ex.4}
Next, we consider the same random $\mathtt{ETA}$ string from the previous example, but this time, we approach the original goal of minimizing $2L+E$. 
For this objective, the surprise of each $\mathtt{A}$ in the target document would incur an encoding cost of $2\cdot\frac{5.6439}{8}$, so it would be cheaper to just make an error. 
As such, we are inclined to only encode the $\mathtt{E}$s and $\mathtt{T}$s in our hints file. 
If we discard the $\mathtt{A}$s from the target document, then we get a random string of $\mathtt{E}$s and $\mathtt{T}$s, which we can encode using the technique in Example~\ref{ex.2}. 
(Notably, the arithmetic coding approach in Example~\ref{ex.3} would deliver an identical hints file.) 
How can we use this hints file at the decoding stage? 
For example, suppose the target document is $\mathtt{ETATEETTT}$. 
Then we discard the $\mathtt{A}$ and encode the remaining string as a one-byte file of hints, namely, $01100111$. 
Recall that at the decoding stage, we receive one character at a time. 
Using the hints file, we first guess $\mathtt{E}$, and then we are told that the correct character is $\mathtt{E}$. 
Next, we read off the 1 to guess $\mathtt{T}$, and then we receive $\mathtt{T}$. 
The third bit in our hints file is 1, so we guess $\mathtt{T}$ next, but then we receive $\mathtt{A}$, meaning our guess was wrong. 
However, we know from our encoding that the next non-$\mathtt{A}$ is necessarily a $\mathtt{T}$, so for our fourth guess, we guess $\mathtt{T}$ again, and this time, we're correct.
\end{example}

\begin{example}
\label{ex.5}
Let's now consider target documents in which characters are allowed to be \textit{statistically dependent}. 
This is an important modeling choice since, for example, you will never see a single letter appearing three times in a row in a standard English word. 
For simplicity, suppose our string is a concatenation of random blocks of characters, each drawn independently from the distribution
\[
\left\{\begin{array}{cl}\mathtt{E}&\text{with probability }49\%\\\mathtt{T}&\text{with probability }49\%\\\mathtt{AS}&\text{with probability }\phantom{0}1\%\\\mathtt{AH}&\text{with probability }\phantom{0}1\%\end{array}\right..
\]
Alternatively, this random process can be represented at the character level in terms of the following two-state \textit{Markov chain}:
\begin{center}
\begin{tikzpicture}[->, >=stealth', auto, semithick, node distance=6cm]
\tikzstyle{every state}=[fill=lightgray,draw=black,thick,text=black,scale=1]
\node[state] (0) {};
\node[state] (1) [right of=0] {};
\node (s) [left of=0, node distance=2cm] {start};
\path
(s) edge (0)
(0) edge[loop above] node{$49\%~\mathtt{E}$} (0)
(0) edge[loop below] node{$49\%~\mathtt{T}$} (0)
(0) edge node{$2\%~\mathtt{A}$}	(1)
(1) edge[bend right,above] node[above]{$50\%~\mathtt{S}$} (0)
(1) edge[bend left,below] node{$50\%~\mathtt{H}$} (0);
\end{tikzpicture}
\end{center}
Suppose we seek to minimize $2L+E$ for a random string that was drawn from this process. 
Then just like in Example~\ref{ex.4}, any appearance of $\mathtt{A}$ comes with a surprise of 5.6438 bits, which is more costly than just making an error. 
As such, we are again inclined to discard any $\mathtt{A}$s from the target document and encode the remaining characters. 
For example, suppose the target document is $\mathtt{ETAHTETTT}$. Then we discard the $\mathtt{A}$ and encode $\mathtt{E}$s and $\mathtt{S}$s with 0s and the $\mathtt{T}$s and $\mathtt{H}$s with 1s, resulting in the byte $01110111$. 
Then we iteratively decode this hints file as $\mathtt{ET\textcolor{red}{T}HTETTT}$, where the red character is an error. 
Notably, after incorrectly interpreting the third bit from the hints file as encoding a $\mathtt{T}$, we inferred a change of state in our Markov chain and correctly reinterpreted the bit in this new context as encoding an $\mathtt{H}$.
\end{example}

\section*{Theory and practice}

Depending on the Markov chain we use to model the target document, we will frequently spend at least a couple of bits to encode a given character. 
That means that at the decode step, when we realize we made an error, we might need to reinterpret more than just the most recent bit from the hints file. 
We refer to this decoding trick as \textit{the rewind mechanism} (a reference to how your ancestors interacted with analog tapes in the 1900s). 
While the discussion in Example~\ref{ex.4} hinted at how to determine which characters we should encode for the rewind mechanism, the optimal choice in general is the subject of a little theorem that we proved specifically for this challenge:

\begin{theorem}
\label{thm.main result}
Take $\alpha\approx0.1854$ such that $(1+\alpha)^{1+\alpha} = (16\alpha)^\alpha$, and consider the largest $k$ such that the $k$th most common character is at least $\alpha$ times as likely as the $k-1$ most common characters combined.
Then encoding the $k$ most common characters minimizes the objective $2L+E$ for the rewind mechanism.
\end{theorem}

(The astute reader may have been tipped off by the weasel words ``for the rewind mechanism'' in the above theorem statement. 
It turns out that the rewind mechanism produces a slightly suboptimal encoding strategy for the objective $2L+E$, but a complicated modification of this approach turns out to be optimal; this will be the subject of a sequel paper. 
In the meantime, the reader is encouraged to find ways to improve upon the rewind mechanism.)

\begin{proof}[Proof sketch of Theorem~\ref{thm.main result}]
The cost of encoding a subset $S$ of characters is
\[
\operatorname{cost}(S)
:=2\cdot\frac{1}{8} \sum_{i\in S}p(i)\log_2\left(\frac{p(S)}{p(i)}\right) + 1-p(S),
\]
where $p(i)$ denotes the probability of character $i$, and $p(S):=\sum_{i\in S}p(i)$.
We start by proving an intermediate claim that for every minimizer $T$ of $\operatorname{cost}(\cdot)$, it holds that $T$ is nonempty and consists of the $|T|$ most common characters.

First, $T$ is nonempty since $\operatorname{cost}(\varnothing)=1$ and $\operatorname{cost}(\{j\})=1-p(j)$ for all $j$.
(In fact, this also establishes our intermediate claim in the case where $|T|=1$.)
The rest of our proof makes use of a certain auxilary function.
Given $S$ and $j\not\in S$, then taking $x:=\frac{p(j)}{p(S)}$, one may verify that
\[
\frac{\operatorname{cost}(S\cup\{j\})-\operatorname{cost}(S)}{p(S)}
=\frac{1}{4}\Big((1+x)\log_2(1+x)-x\log_2x\Big)-x
=:f(x).
\]
Furthermore, it holds that
\begin{itemize}
\item[(i)]
$f(x)$ is positive for every $x\in (0,\alpha)$,
\item[(ii)]
$f(\alpha)=0$ and $f(x)$ is negative for every $x\in(\alpha,\infty)$, and
\item[(iii)]
$f(x)$ is decreasing at every $x\in[\alpha,\infty)$.
\end{itemize}
Take a minimizer $T$ of $\operatorname{cost}(\cdot)$ with $|T|\geq2$, consider its least likely character $j\in T$, and put $S:=T\setminus\{j\}$.
Then $p(j)\geq\alpha p(S)$, since otherwise $\operatorname{cost}(S)<\operatorname{cost}(T)$ by~(i).
Also, there is no $j'\not\in T$ with $p(j')> p(j)$, since otherwise $\operatorname{cost}(S\cup\{j'\})<\operatorname{cost}(T)$ by~(iii).
Thus, $T$ consists of the $|T|$ most common characters, as claimed.

To conclude the proof of our original result, take any minimizer $T$ of $\operatorname{cost}(\cdot)$ that is maximal as a set (and with any cardinality), and consider the most likely character $j''\not\in T$.
Then by (ii), our assumption that $\operatorname{cost}(T\cup\{j''\})>\operatorname{cost}(T)$ implies $p(j'')<\alpha p(T)$, and so $|T|=k$.
\end{proof}

After training a Markov model that best predicts characters in the \textit{Moby-Dick} corpus, we applied Theorem~\ref{thm.main result} to inform an encoding scheme using the rewind mechanism, and the result was the lowest scoring submission at the time. 
As an illustration of this method's performance, let's take a look at how it might decode the first line of \textit{Moby-Dick} using a Markov model that was trained on the remainder of the book:

\bigskip

\footnotesize

\noindent
\begin{Verbatim}[commandchars=\\\{\}]
Call me Ishmael. Some years ago--never mind how long precisely--having little or no
C\hl{Hp}l \hl{a}e \hl{a}\hl{  }\hl{  }mael. \hl{C}ome \hl{s}ears ago\hl{  }-n\hl{o}ver mind how \hl{h}ong \hl{s}resisely\hl{  }-\hl{were}ng \hl{t}ittle \hl{b}r no

money in my purse, and nothing particular to interest me on shore, I thought I would
\hl{a}o\hl{r}\hl{  }y in \hl{t}y \hl{ha}rse, and \hl{l}othing \hl{s}articular to \hl{p}n\hl{  }erest \hl{t}e on \hl{t}hore\hl{  } \hl{t} \hl{s}hought I would

sail about a little and see the watery part of the world.
\hl{te}il about \hl{t} little \hl{i}nd \hl{l}ee the w\hl{h}ter\hl{  } part of the w\hl{a}rld\hl{  }
\end{Verbatim}

\normalsize

\bigskip

Since this passage appears after a few hard returns in the book, the model initially guesses that the text will be \texttt{CHAPTER}, and in fact, it is correct for the first character. 
However, once it sees the \texttt{a} of \texttt{Call}, it begins to think that the first word is \texttt{Captain} (\`{a} la \texttt{Captain Ahab}). 
After learning that this guess is also wrong, it finally decides that the intended word is \texttt{Call}. 
Interestingly, the decoder didn't recognize \texttt{Ishmael} until seeing the first three characters.
Later, after seeing the \texttt{w} in \texttt{watery}, the model expected the word to be \texttt{white}, as in \texttt{white whale}.

\section*{Concluding remarks}

We used ideas from information theory to predict the text of \textit{Moby-Dick} one character at a time.
To achieve better performance in this task, one can try to improve the rewind mechanism or the underlying Markov model. 
For an alternative to the rewind mechanism that achieves the best possible performance for a given model, see our forthcoming sequel paper.
Meanwhile, one might find a better Markov model using some relevant ideas behind modern large language models~\cite{Sanderson:online}.
We encourage you to try your hand at chasing this whale!

\end{document}